\documentclass[12pt]{article}
\usepackage{ametsoc}

\newcommand{\myabstract}{We offer a new method for determining the wind source term for energy and momentum fluxes transfer from the atmosphere to the wind-driven sea. This new source-term formulation is based on extensive analysis of experimental data collected at different sites around the world. It is shown that this  new wind source term to be consistent both with numerical solution of exact equation for resonant four-wave interactions and available experimental data.}

\begin{document}

\title{\textbf{\large{New wind input term consistent with experimental, \\ theoretical and numerical considerations}}}

\author{
\textsc{Zakharov V.E.}\\
\centerline{\textit{\footnotesize{Department of Mathematics, University of Arizona, Tucson, AZ 85721, USA}}}\\
\centerline{\textit{\footnotesize{Novosibirsk State University, Novosibirsk, 630090, Russia}}}\\
\centerline{\textit{\footnotesize{Lebedev Physical Institute RAS, Leninsky 53, Moscow 119991, Russia}}}\\
\centerline{\textit{\footnotesize{Waves and Solitons LLC, 1719 W. Marlette Ave., Phoenix, AZ 85015, USA}}}\\
\textsc{Resio D.}\\
\centerline{\textit{\footnotesize{Taylor Engineering Research Institute, University of North Florida}}}\\
\centerline{\textit{\footnotesize{1 UNF Drive, Jacksonville, FL 32224, USA}}}\\
\textsc{Pushkarev A.}\\
\centerline{\textit{\footnotesize{Waves and Solitons LLC, 1719 W. Marlette Ave., Phoenix, AZ 85015, USA}}}\\
\centerline{\textit{\footnotesize{Novosibirsk State University, Novosibirsk, 630090, Russia}}}\\
\centerline{\textit{\footnotesize{Lebedev Physical Institute RAS, Leninsky 53, Moscow 119991, Russia}}}
}

\ifthenelse{\boolean{dc}}
{
\twocolumn[
\begin{@twocolumnfalse}
\amstitle

\begin{center}
\begin{minipage}{13.0cm}
\begin{abstract}
\myabstract
	\newline
	\begin{center}
		\rule{38mm}{0.2mm}
	\end{center}
\end{abstract}
\end{minipage}
\end{center}
\end{@twocolumnfalse}
]
}
{
\amstitle
\begin{abstract}
\myabstract
\end{abstract}
\newpage
}

\section{Introduction}

Presently, most oceanographers believe that a satisfactory description of the the evolution of wave action in a wind-driven sea is given by  (see \cite{Z1}, \cite{BPRZ} and \cite{BBRZ}):
\begin{eqnarray}
\label{GenEq}
\frac{d N_{{\bf k}}}{d t} = \frac{\partial N_{{\bf k}}}{\partial t} + \frac{\partial \omega_k}{\partial {\bf k}} \frac{\partial N_{{\bf k}}}{\partial {\bf r}} = S_{nl}+S_s
\end{eqnarray}
where $N=N({\bf k},{\bf r},t)$ is the wave action density, ${\bf k}=(k_x,k_y)$ is the Fourier space wave-number vector, ${\bf r}=(x,y)$ is the real-space coordinate vector, $\omega_k=\sqrt{gk}$ (where $k=|{\bf k}|$) is the dispersion law for gravity waves, $S_{nl}$ is the source term for resonant four-wave interactions, and $S_s = S_{in}+S_{diss}$ is the source term responsible for energy transfer from the wind and dissipation due to wave breaking and other forms of dissipation within the wave field.
 
The $S_{nl}$ term is a nonlinear integral operator which, in a global sense, can be written as a ”conservative” kinetic equation 
\begin{eqnarray}
\label{HasEq}
\frac{d N_{\bf k}}{dt} = S_{nl}
\end{eqnarray}
This form conserves total wave action and formally conserves energy and momentum; however, conservation is not realized over a finite range of frequencies (see \cite{PZ}). The associated nonlinear fluxes have been shown to produce power Kolmogorov-like tails in a finite time, transferring energy and momentum into the area of small wave numbers. The first spatially homogeneous isotropic solution of that type was found by \cite{ZF}
\begin{eqnarray}
\label{KZspectrum}
\epsilon(\omega) = C_1 \frac{P_0^{1/3}}{\omega^4}
\end{eqnarray}
where spectral energy density $\epsilon(\omega,\theta)$ ($\theta$ is the wave-number vector angle) is connected with wave action density $N({\bf k},t)$ by the relation
\begin{eqnarray}
\label{RelationEpsilonN}
\epsilon(\omega,\theta)d\omega d\theta=\omega_k N_{{\bf k}} d{\bf k}
\end{eqnarray}
$P_0$ is the energy flux to high wave numbers and $C_1$ is the Kolmogorov constant $C_1 \simeq 4 \pi \cdot 0.219 = 2.75$ (see \cite{Z}).

In contrast to the strong theoretical basis for $S_{nl}$, our knowledge of the wind input and wave dissipation source terms $S_s$ is quite poor. The creation of a reliable, well justified theory for wind input $S_in$ has been hindered by the presence of strong turbulence in the air boundary layer over the sea surface. Even the one of the most crucial elements for this theory, the vertical profile of the mean horizontal wind velocity, is still subject to debate in the region closest to the ocean surface where wave motions interact strongly with atmospheric motions. 

Direct measurements of $S_{in}$ are scarce. As a result, we currently have many different heuristic models of $S_{in}$ and the scatter between different models is large in terms of both the predicted total fluxes and the distribution of fluxes into different frequencies and directions. For instance, \cite{DP} form for $S_{in}$ predicts magnitudes of $S_{in}$ approximately five times higher then the \cite{HS} form for $S_{in}$. A comparison of different forms for of $S_{in}$ is presented by \cite{BPRZ}.

Unfortunately, but our understanding and quantification of the dissipation source term $S_{diss}$ is not any better. The theory is not well developed and experimental data are far from complete; consequently, the forms for $S_{diss}$ used in operational models are heuristic and not well justified.
 
The situation in wave models is aggravated by the fact that the heuristic forms for wind input and wave dissipation are of the same magnitude as the nonlinear source term in essentially all parts of the wave spectrum, i.e. $S_s \sim S_{nl}$. This viewpoint was postulated by \cite{P} and since this time has been commonly accepted. Some authors, for example, \cite{DHH}, \cite{DSTH}) even assume that $S_{nl}$ is negligibly small with respect to $S_s$. According to this opinion, the wind-generation process at full development is arrested by wave-breaking; although it is difficult to find strong theoretical or empirical support for these hypotheses.

Although the magnitude of $S_{nl}$ can be small in regions of a spectrum exhibiting constant fluxes, in many situations the nonlinear term $S_{nl}$ surpasses the $S_s$ source term by at least the order of magnitude in many regions of the spectrum. It has been argued by \cite{Z}, \cite{ZB} that this makes the role of $S_{nl}$ absolutely critical to the evolution of wind wave spectra. In the next three sections, we shall utilize experimental evidence, theoretical considerations and numerical simulations to formulate and test a new wind input source term that functions in concert with $S_{nl}$ to produce wave growth and wave spectra in agreement with the findings of Badulin et al. (2005, 2007).

\section{Experimental evidence}

We start here by examining empirical evidence from around the world which has been utilized to quantify energy levels within the equilibrium range in spectra by \cite{RL}.  For convenience, we shall also use the same notation used by \cite{RL} in their study, for the directionally integrated spectral energy densities in frequency and wavenumber space,
\begin{eqnarray}
\label{ZKspec2}
E_4(f) &=& \frac{2\pi \alpha_4 V g}{(2 \pi f)^4} \\
\label{ZKspec3}
F_4(k) &=& \beta k^{-5/2} 
\end{eqnarray}
where $f=\frac{\omega}{2 \pi}$, $\alpha_4$ is the constant, $V$ is some characteristic velocity and $\beta=\frac{1}{2}\alpha_4 V g^{-1/2}$. These notations are based on relation of spectral densities $E(f)$ and $F(k)$ in frequency $f=\frac{\omega}{2\pi}$ and wave-number ${\bf k}$ basises:
\begin{eqnarray}
\label{EnRel}
F(k)=\frac{c_g}{2\pi}E(f)
\end{eqnarray}
where $c_g=\frac{d \omega}{d k}$ is the group velocity.

\begin{figure}[t]
\noindent\center\includegraphics[scale=1.4]{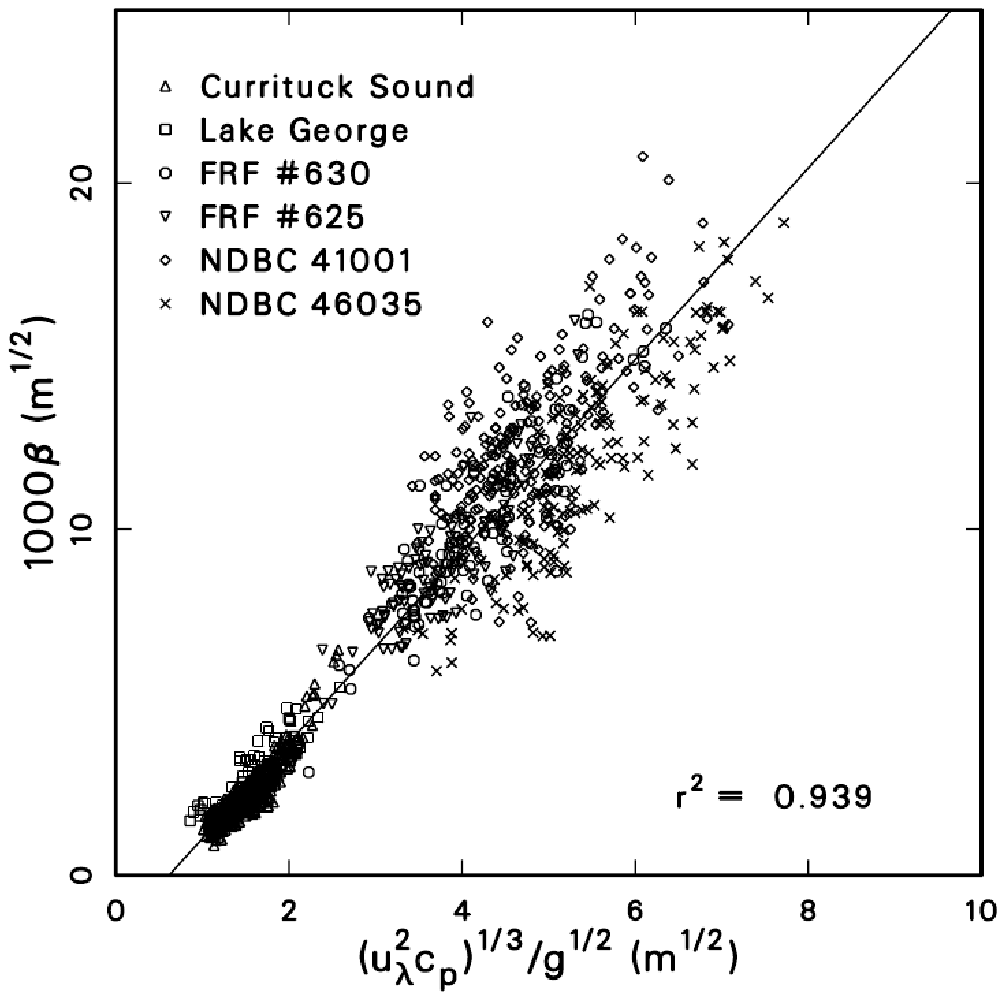} 
\caption{{\it Correlation of equilibrium range coefficient $\beta$ with $(u_\lambda^2 c_p)^{1/3}/g^{1/2}$ based on data from six disparate sources. Adopted from \cite{RL}}} \label{RegLine}
\end{figure}

The notations in Eqs.(\ref{ZKspec2})-(\ref{ZKspec3}) are connected with the spectral energy density $\epsilon(\omega,\theta)$ (see Eqs.(\ref{KZspectrum})-(\ref{RelationEpsilonN})) through
\begin{eqnarray}
E(f) = 2 \pi \int \epsilon(\omega,\theta) d\theta
\end{eqnarray}
The \cite{RL} analysis showed that experimental energy spectra $F(k)$, estimated through averaging $<k^{5/2}F(k)>$, can be approximated by linear regression line as a function of $(u^2_\lambda c_p)^{1/3} g^{-1/2}$. Fig.\ref{RegLine} shows that the regression line 
\begin{eqnarray}
\beta=\frac{1}{2} \alpha_4 \left[ (u_\lambda^2 c_p)^{1/3}-u_0\right]g^{-1/2}
\end{eqnarray}
indeed, seems to be a reasonable approximation of these observations.
Here $\alpha_4=0.00553$, $u_0=1.93$ $m/sec$, $c_p$ is the spectral peak phase speed and $u_\lambda$ is the wind speed at the elevation equal to a  fixed fraction  $\lambda=0.065$ of the spectral peak wavelength $2\pi/k_p$, where $k_p$ is the spectral peak wave number. \cite{RL} assumed that the near surface boundary layer can be treated as neutral and thus follow a conventional logarithmic profile   
\begin{eqnarray}
\label{LogProfile}
u_\lambda=\frac{u_\star}{\kappa} \ln \frac{z}{z_0}
\end{eqnarray}
having  Von Karman coefficient $\kappa=0.41$, where $z=\lambda \cdot 2\pi/k_p$ is the elevation equal to a fixed fraction $\lambda = 0.065$ of the spectral peak wavelength $2\pi/k_p$,  where $k_p$ is the spectral peak wave number, and $z_0=\alpha_C u_{\star}^2 /g$ subject to \cite{C} surface roughness with $\alpha_C = 0.015$.

\section{Theoretical considerations}
Self-similar solutions of conservative kinetic equation (\ref{HasEq}) were studied in \cite{Z1}, \cite{BPRZ} and \cite{BBRZ}. In this chapter we study self-similar solutions of the forced kinetic equation
\begin{eqnarray}
\label{ForcedHasEq}
\frac{\partial \epsilon(\omega,\theta)}{\partial t} = S_{nl}+\gamma(\omega,\theta) \epsilon(\omega,\theta)
\end{eqnarray}
where $\epsilon(\omega,\theta)=\frac{2\omega^4}{g}N({\bf k},\theta)$ is the energy spectrum. For our purposes, it is sufficient to simply use the dimensional estimate for $S_{nl}$,
\begin{eqnarray}
S_{nl} \simeq \omega \left( \frac{\omega^5 \epsilon}{g^2} \right)^2 \epsilon
\end{eqnarray}
Eq.(\ref{ForcedHasEq}) has a self-similar solution if
\begin{eqnarray}
\gamma(\omega,\theta) = \alpha \omega^{1+s} f(\theta)
\end{eqnarray}
where $s$ is a constant. Looking for self-similar solution in the form 
\begin{eqnarray}
\epsilon(\omega,t) = t^{p+q} F(\omega t^q)
\end{eqnarray}
we find
\begin{eqnarray}
q&=&\frac{1}{s+1} \\
p&=&\frac{9 q-1}{2} = \frac{8-s}{2 (s+1)}
\end{eqnarray}
The function $F(\xi)$ has the maximum at $\xi\sim\xi_p$, thus the frequency of the spectral peak is 
\begin{equation}
\omega_p\simeq\xi_p t^{-q}
\end{equation}
The phase velocity at the spectral peak is
\begin{equation}
c_p = \frac{g}{\omega_p} = \frac{g}{\xi_p} t^q = \frac{g}{\xi_p} t^{\frac{1}{s+1}}
\end{equation}
According to experimental data, the main energy input into the spectrum occurs in the vicinity of the spectral peak, i.e. at $\omega\simeq \omega_p$. For $\omega>>\omega_p$, the spectrum is described by Zakharov-Filonenko tail
\begin{equation}
\epsilon(\omega) \sim P^{1/3} \omega^{-4}
\end{equation}
Here 
\begin{equation}
P = \int_0^\infty \int_0^{2 \pi} \gamma(\omega,\theta) \epsilon(\omega,\theta) d\theta
\end{equation}
This integral converge if $s<2$. For large $\omega$ 
\begin{eqnarray}
\epsilon(\omega,t) \simeq \frac{t^{p-3q}}{\omega^4} \simeq \frac{t^{\frac{2-s}{2(s+1)}}}{\omega^4}
\end{eqnarray}
More accurately
\begin{eqnarray}
\label{ZKspec1} 
\epsilon(\omega,t) &\simeq& \frac{\mu g}{\omega^4} u^{1-\eta} c_p^\eta g(\theta) \\
\eta &=& \frac{2-s}{2}
\end{eqnarray}
Now supposing $s=4/3$ and $\gamma\simeq \omega^{7/3}$, we get $\eta=1/3$ , which is exactly experimental regression line prediction. Because it is known from regression line on Fig.\ref{RegLine} that $\xi=1/3$, we immediately get $s=4/3$ and the wind input term
\begin{eqnarray}
\label{NewWindInput}
S_{wind} \simeq \omega^{7/3} \epsilon
\end{eqnarray}
One should note that the dependence shown in Eq. (\ref{NewWindInput}) has already been predicted by \cite{RP} from dimensional consideration.

\section{Numerical simulations}
To check the self-similar conjecture implicit in our formulation shown in Eq.(\ref{NewWindInput}) we performed a series of numerical simulations of Eq.(\ref{GenEq}) in the time domain ($\frac{\partial N}{\partial {\bf r}}=0$, or spatially homogeneous) and fetch limited  ($\frac{\partial N}{\partial t} =0$, or spatially inhomogeneous) situations. The same input term Eq.(\ref{NewWindInput}) has been used in both cases in the form
\begin{eqnarray}
\label{Has2}
S_{wind} &=& 0.2 \frac{\rho_{air}}{\rho_{water}} \omega \left( \frac{\omega}{\omega_0}\right)^{4/3} f(\theta), \nonumber
\\
f(\theta) &=&	\left\{ \begin{array}{rcl}
		\cos{\theta} & \mbox{for} & -\pi/2 \leq \theta \leq \pi/2 \\
		0 & \mbox{otherwise} 
		\end{array} \right. \nonumber
\\
\omega_0 &=& \frac{g}{u},\,\,\, \frac{\rho_{air}}{\rho_{water}}=1.3\cdot 10^{-3} \nonumber
\end{eqnarray}
where $u$ is the wind speed, $\rho_{air}$ and $\rho_{water}$ are the air and water density correspondingly. 

Wind speed $u$ is taken here as the speed at a reference level of 10 meters ($u_{10}$). To make comparison with experimental results of \cite{RL}, we used relation $u_{\star} \simeq u_{10}/28$ (see \cite{G}) in Eq.(\ref{LogProfile}).

Both situations also need knowledge of the dissipation term $S_{diss}$, which is taken into account here in the same way, as was proposed by Resio and Long (2007), where white-capping dissipation term $S_{diss}$ was introduced implicitly through an $f^{-5}$ energy spectral tail stretching in frequency range from $f_d =1.1$ to $f_{max} = 2.0$. To date, this approach has been shown by both experimental observations and numerical experiments (see \cite{RL}) to provide an effective sink for energy in direct cascades into high frequencies.

\subsection{Time domain simulations}
\begin{figure}[t]
\noindent\center\includegraphics[scale=1.]{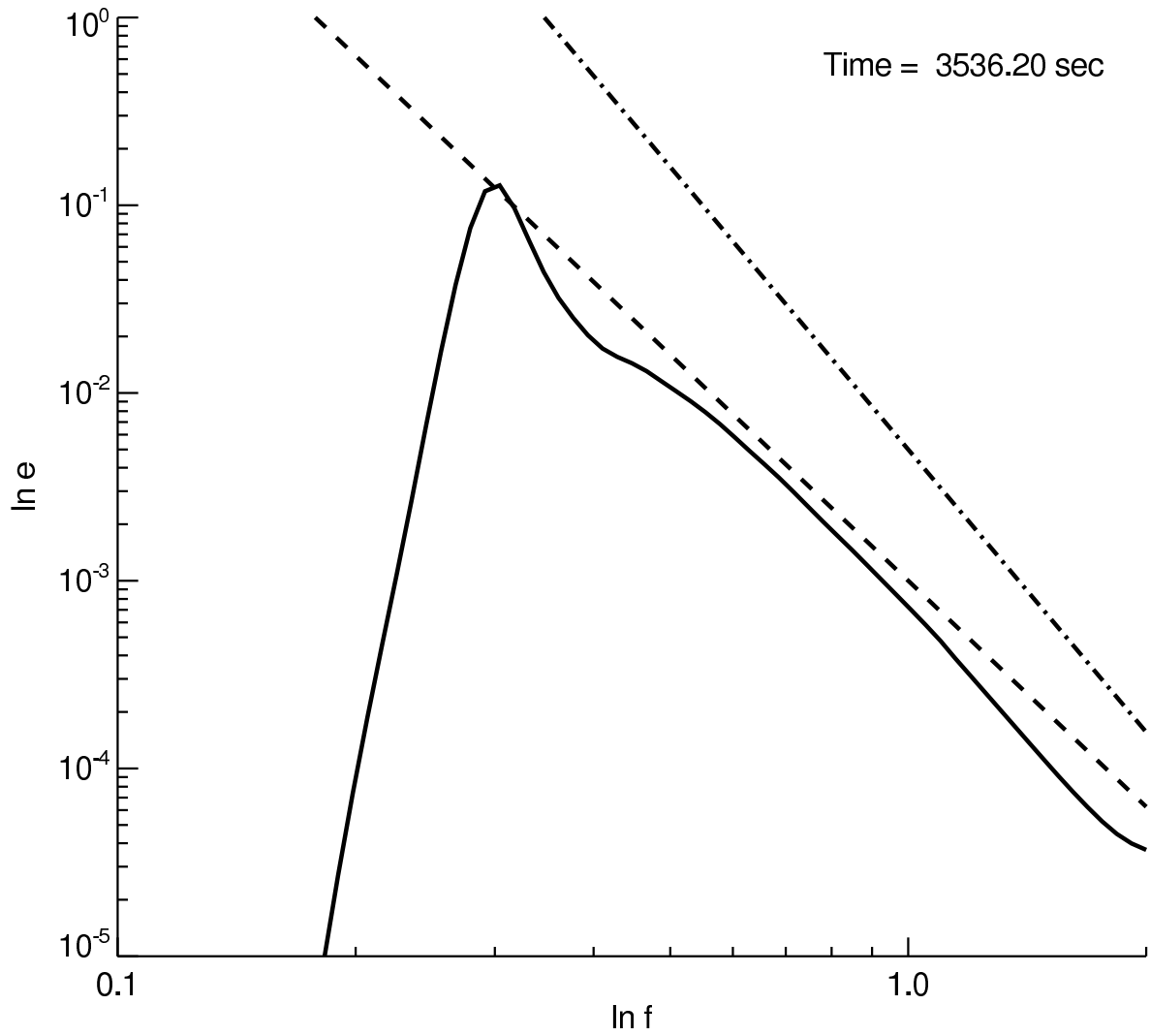} 
\caption{{\it Typical picture of $\ln{\epsilon (f)}$ as a function of $\ln{f}$, wind speed $u=10.0$ m/sec. Solid line -- directional spectrum, dashed line -- spectrum $f^{-4}$, dash-dotted line -- spectrum $f^{-5}$.}} \label{Spectrum}
\end{figure}

Time domain numerical simulations have been started from an initial uniform noise energy distribution in Fourier space. The typical picture of the directional energy spectrum for the wind speed $u = 10 m/sec$ is presented on Fig.\ref{Spectrum}. In this figure, one can distinguish three separate areas of the spectrum – area of spectral peak, intermediate portion of the spectral tail proportional to $f^{-4}$ and high-frequency portion of the spectrum, proportional to $f^{-5}$.

\begin{figure}[t]
\noindent\center\includegraphics[scale=1.]{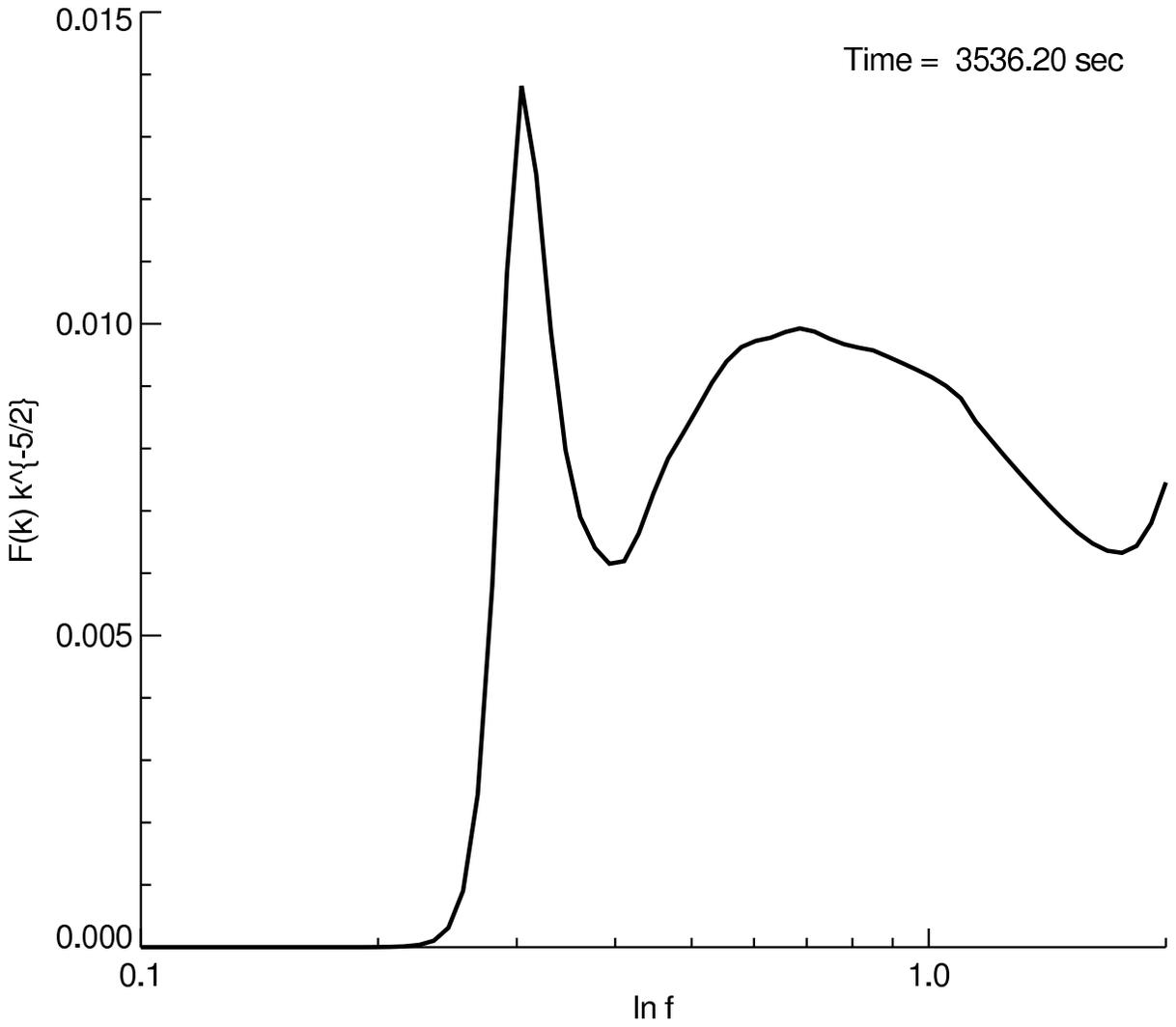}
\caption{{\it Compensated spectrum $F(k) k^{5/2}$ as a function of $\ln{f}$,  wind speed $u=10.0$ m/sec}} \label{Beta}
\end{figure}

The compensated spectrum $F(k) \cdot k^{5/2}$ from these simulations is presented on Fig.\ref{Beta}. One can see plateu-like region responsible for $k^{-5/2}$ behavior, equivalent to $f^{-4}$ tail in Fig.\ref{Spectrum}. This exact solution of Eq.(\ref{GenEq}) was found by \cite{ZF}.

Now we are turning to direct comparison of the numerical simulation  with the experimental analysis by \cite{RL} presented on Fig.\ref{RegLine}.

\begin{figure}[t]
\noindent\center\includegraphics[scale=0.8]{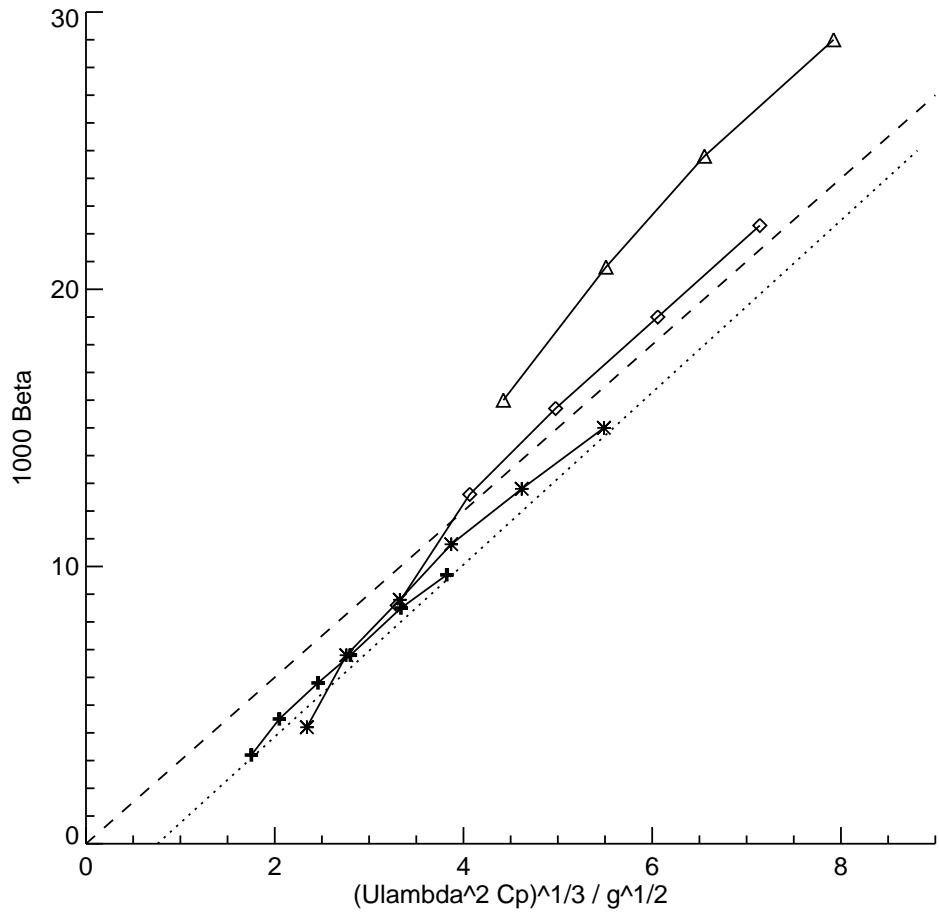} \\
\caption{{\it Experimental, theoretical and numerical evidence on the single graph for $1000 \beta$ as a function of $(u_\lambda^2 c_p)^{1/3}/g^{1/2}$. Experimental result: dotted line -- experimental regression line from Fig.\ref{RegLine}. Theoretical result: dashed line -- theoretical relation Eq.(\ref{TheorRegr}). Numerical results: crosses correspond to  $u=2.5$, stars to $u=5.0$, rectangles to $u=10.0$, triangles to $u=20.0$}}\label{NewForcing}
\end{figure}

Fig.\ref{NewForcing} presents the plot of the function $\beta=F(k)\cdot k^{5/2}$ in terms of $(u_{\lambda}^2 C_p)^{1/3}/g^{1/2}$ for four different runs, corresponding to wind speeds $u=2.5, 5.0, 10.0, 20.0$ $m/sec$, along with the regression line from \cite{RL}. The overall correspondence with the regression line for values of $u=2.5, 5.0, 10.0$ $m/sec$ is quite good. The results corresponding to $u=20.0$ $m/sec$ are bit off of the regression line, but exhibit the same slope.  

Another important theoretical relationship can be derived from joint consideration of Eqs.  (\ref{ZKspec2}), (\ref{EnRel}) and (\ref{ZKspec1}):
\begin{eqnarray}
\label{TheorRegr}
1000\beta = 3 \frac{(u^2 c_p)^{1/3}}{g^{1/2}}
\end{eqnarray}
This dependence, presented on  Fig.\ref{NewForcing}, shows that using this new wind input term produces a good correspondence of theory, experiment and numerical results.

\subsection{Limited fetch numerical simulations}
A limited fetch equivalence to the time-domain situation can be described by reduction of Eq.(\ref{GenEq})
\begin{eqnarray}
\label{LFeq}
\frac{1}{2} \frac{g \cos \theta}{\omega } \frac{\partial \epsilon}{\partial x} = S_{nl}(\epsilon)+S_{wind}+S_{diss}
\end{eqnarray}
where ${\bf x}$ is the coordinate orthogonal to the shore and $\theta$ is the angle between individual wavenumber ${\bf k}$ and the axis ${\bf x}$.

Eq.(\ref{LFeq}) is somewhat difficult for nimerical simulation, since it contains a singularity in the form of $\cos \theta$ in front of $\frac{\partial \epsilon}{\partial x}$. We overcame this problem of division by zero through zeroing one half of the Fourier space of the system for the waves propagating toward the shore. Since it is well-- known that the energy in such waves is small with respect to waves propagating in the offshore direction, such approximation is quite reasonable for our purposes.

The same sort  of self-similar analysis performed for time domain situation can be also repeated for limited fetch one. The result of inserting the self-similar substitution 
\begin{eqnarray}
\label{SSSS}
\epsilon = x^{p+q}F(\omega x^{q})
\end{eqnarray}
into Eq.(\ref{LFeq}) gives the values of indices
\begin{eqnarray}
p=\frac{10q-1}{2},\,\,\,\,q=\frac{1}{s+2}
\end{eqnarray}
and
\begin{eqnarray}
q=\frac{3}{10},\,\,\,\,p=1,\,\,\,\,s=\frac{4}{3}
\end{eqnarray}
We find that in the fetch-limited case the  wind forcing index $s$ is similar to time domain case when the wind forcing is given by Eq.(\ref{NewWindInput}). Therefore, numerical simulation of Eq.(\ref{LFeq}) has been performed for the same input functions as in the time doman case with the same initial conditions for the form  of low-level energy noise in Fourier space.

\begin{figure}[t]
\noindent\center\includegraphics[scale=0.9]{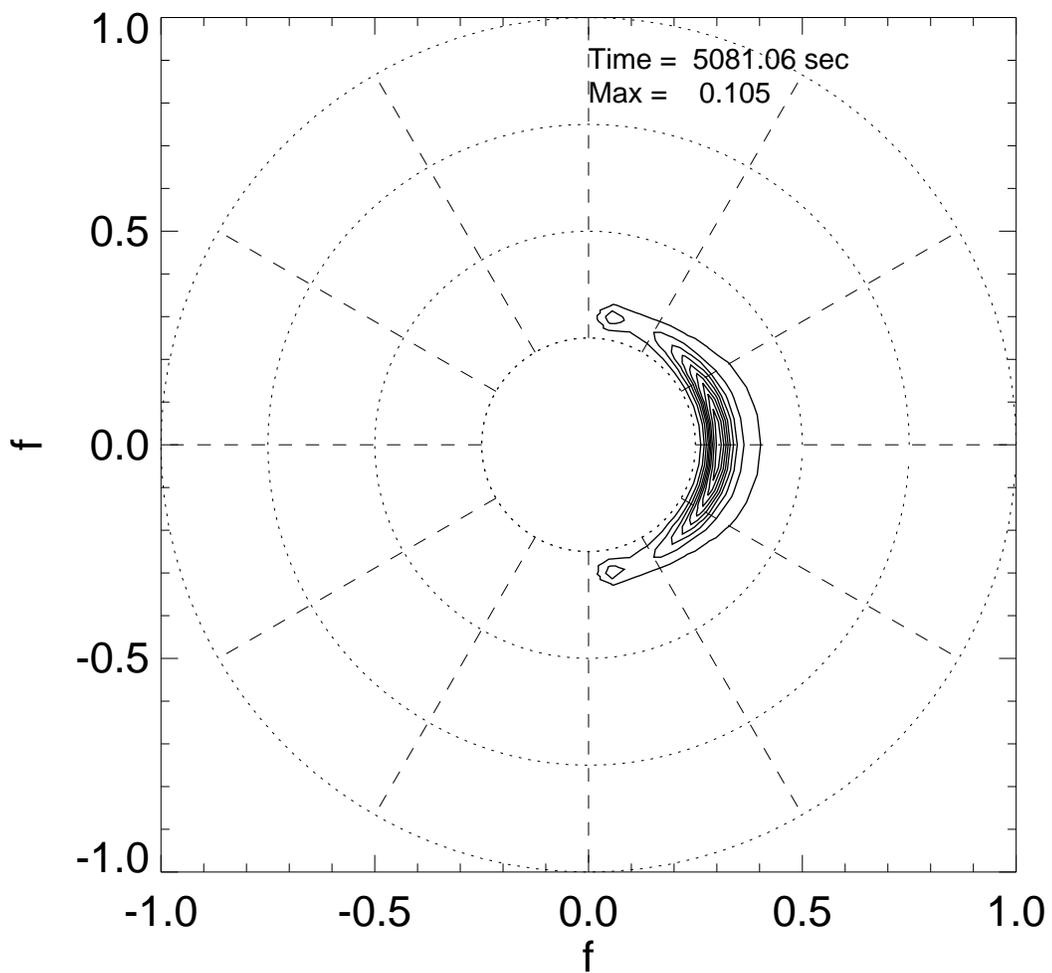} 
\caption{{\it Energy spectral density $\epsilon(f,\theta)$ as a function of frequency $f$ and angle $\theta$ in polar coordinates.}} \label{Energy3D}
\end{figure}

\begin{figure}[t]
\noindent\center\includegraphics[scale=0.8]{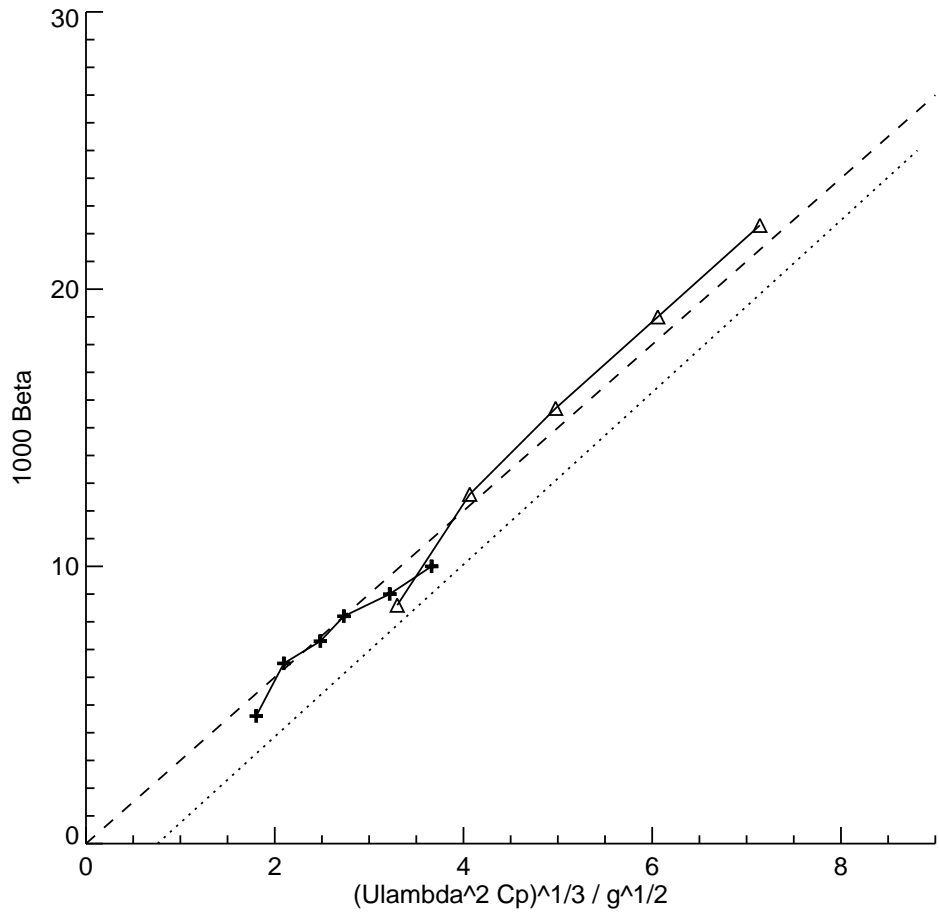} 
\caption{{\it Two simulations for wind input $10m/sec$ : time-limited domain (triangles) and limited fetch (crosses).  Dotted line -- correlation of the equilibrium range coefficient $\beta$ with $(u_\lambda^2 c_p)^{1/3}/g^{1/2}$ based on data from six disparate sources adopted from Resio at al, 2004. Dashed line -- theoretical value of equilibrium range coefficient $\beta$.}} \label{RegLineLF}
\end{figure}

Fig.\ref{Energy3D} presents characteristic spectral energy distribution for our limited fetch simulations. Fig.\ref{RegLineLF} presents the plot of function $\beta=F(k)\cdot k^{5/2}$ in terms of the parameter $(u_{\lambda}^2 C_p)^{1/3}/g^{1/2}$ for two different runs, corresponding to the same wind speed $u=10.0$ $m/sec$: one for time domain simulations and another for limited fetch simulations. Both sets of simulations show good correspondence with the \cite{RL} regression line.

\section{Conclusion}
We have inroduced a new form for wond input, similar to that derived from dimensional considerations by \cite{RP}. Numerical simulations using this new source term show good agreement with observed equilibrium range characteristics around the world and produce spectral shapes and energy growth rates which are consistent with the self-similar theoretical arguments of \cite{Z1}, \cite{BPRZ} and \cite{BBRZ}. It is believed that this new form of wind input term can improve the quality of ocean wave prediction in operational models forecasts. 

\begin{acknowledgment} 
This work was sponsored by ONR grant N00014-10-1-0991, NSF grant \# 1130450, Russian Government contract $11.9.34.31.0035$, RFBR grant 12-01-00943, the Program of the RAS Presidium ``Fundamental Problems of Nonlinear Dynamics in Mathematical and Physical Sciences'' and the "Leading scientific schools of Russia" grant NSh 6170.2012.2. Authors gratefully acknowledge continuous support of these foundations.
\end{acknowledgment}

\ifthenelse{\boolean{dc}}
{}
{\clearpage}
\bibliographystyle{ametsoc}
\bibliography{references}

\end{document}